\documentclass[a4paper,twoside]{article}
\usepackage{epsfig}
\usepackage{subfigure}
\usepackage{calc}
\usepackage{amssymb}
\usepackage{amstext}
\usepackage{amsmath}
\usepackage{amsthm}
\usepackage{multicol}
\usepackage{multirow}
\usepackage{pslatex}
\usepackage{apalike}
\usepackage{SCITEPRESS}     
\usepackage{algpseudocode}
\usepackage{algorithm}
\usepackage{xcolor}
\definecolor{xred}{RGB}{255,0,0}
\definecolor{xblue}{RGB}{68,114,196}
\definecolor{xgreen}{RGB}{0,176,80}
\definecolor{xpurple}{RGB}{112,48,160}
\definecolor{xorange}{RGB}{237,125,49}

\makeatletter
\renewcommand{\ALG@name}{Interaction}
\makeatother

\begin{document}
\long\def\/*#1*/{}

\title{ A Framework for Measuring the Costs of Security at Runtime }

\author{
	\authorname{Igor Ivkic\sup{1, 2}, Harald Pichler\sup{2}, Mario Zsilak\sup{2}, Andreas Mauthe\sup{1} and Markus Tauber\sup{2}}
	\affiliation{\sup{1}Lancaster University, Lancaster, UK}
	\affiliation{\sup{2}University of Applied Sciences Burgenland, Eisenstadt, Austria}
	\email{\{i.ivkic, a.mauthe\}@lancaster.ac.uk, \{harald.pichler, 1710635019, markus.tauber\}@fh-burgenland.at}
}

\keywords{Cyber-Physical Systems, Internet of Things, Component Monitoring, Task Tracing, Security Cost Modelling.}

\abstract{In Industry 4.0, Cyber-Physical Systems (CPS) are formed by components, which are interconnected with each other over the Internet of Things (IoT). The resulting capabilities of sensing and affecting the physical world offer a vast range of opportunities, yet, at the same time pose new security challenges. To address these challenges there are various IoT Frameworks, which offer solutions for managing and controlling IoT-components and their interactions. In this regard, providing security for an interaction usually requires performing additional security-related tasks (e.g. authorisation, encryption, etc.) to prevent possible security risks. Research currently focuses more on designing and developing these frameworks and does not satisfactorily provide methodologies for evaluating the resulting costs of providing security. In this paper we propose an initial approach for measuring the resulting costs of providing security for interacting IoT-components by using a Security Cost Modelling Framework. Furthermore, we describe the necessary building blocks of the framework and provide an experimental design showing how it could be used to measure security costs at runtime.}
\onecolumn \maketitle \normalsize \vfill
\section{\uppercase{Introduction}}
\label{sec:introduction}
{\tiny }
\noindent In recent years, cloud computing has changed the way how computer resources are being managed, configured, accessed, and used \cite{ref01}. At the same time, it paved the way towards the fourth industrial revolution (Industry 4.0), which is driven by Cyber-Physical Systems (CPS) and the Internet of Things (IoT) \cite{ref02,ref03}. A CPS is formed by a number of components (e.g. IoT-components), which are interconnected over the IoT and are capable of sensing and affecting the physical world \cite{ref04}. Consequently, the swarm of interacting components tends to quickly become complex and challenging to administer. To address this challenge, various frameworks \cite{ref05} offer solutions to the management of IoT-components entering a CPS \cite{ref06} and to control how they interact with other components in a secure manner. 

Within an interaction a number of components perform different tasks and communicate with each other to serve a specific purpose. To provide security for these interactions, the execution of additional (security-related) tasks, which are not directly linked to the purpose of the interaction, is required. For instance, as shown in Figure 1 the purpose of the inter-

\noindent \begin{figure}[!h]
	\vspace{-1.2cm}
	\centering
	{\epsfig{file = 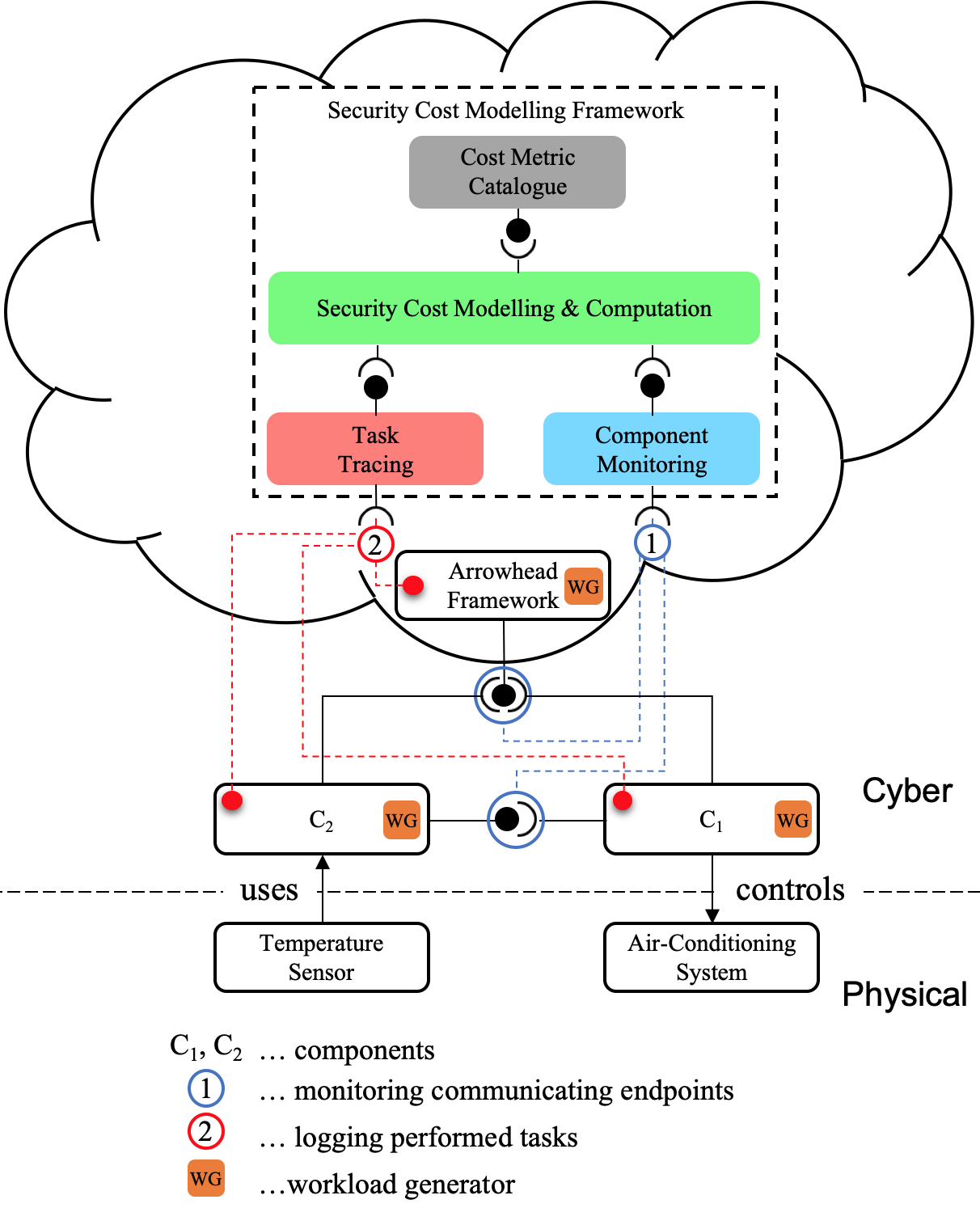, width = \columnwidth}}
	\caption{Security Cost Modelling Framework.}
	\label{fig:fig1}
\end{figure} 

\noindent action is to measure the temperature of a physical room ($C_2$) and to cool it down if necessary ($C_1$). To provide security for this interaction, the IoT Framework ensures that only $C_1$ and $C_2$ are authorised to exchange room temperature data to control the air-conditioning system. Additionally, before any data is transmitted between the components, they encrypt their messages to avoid eavesdropping attacks. These additional steps (authorisation and encryption) are mainly security-related tasks, which are not directly linked to the purpose of the interaction and produce costs (e.g. execution time, computing resources, etc.). 

Due to the complexity of interactions, the number of participating components and performed tasks, an approach is needed to measure how much it costs for providing security at runtime. Measuring security costs of an interaction enables (i) redesigning interactions to produce less security-related costs, (ii) predicting future security costs based on past measurements, and (iii) detecting anomalies based on the expected security costs in comparison to actually measured security costs of an interaction. To address this challenge, we propose an initial approach to automatically measure the resulting costs of providing security by using a Security Cost Modelling Framework as shown in Figure 1. The proposed framework is an extension of our previous work \cite{ref07}, where we proposed an Onion Layer Model, which formally describes a CPS including its interactions, the participating components and their performed tasks. 

In this paper, we extend the Onion Layer Model by proposing a Security Cost Modelling Framework, which uses additional mechanisms to collect data about the interacting components (Component Monitoring) and their performed tasks (Task Tracing). The gathered runtime data is then combined with a Cost Metric Catalogue, which contains security cost metrics and is used to measure the security-related tasks performed by components during interactions. Furthermore, we extend the Onion Layer Model to be able to measure security costs at a specific point in time, explain how these mechanisms could be used in a harness for measuring security costs and discuss how they can be implemented.

The remainder of this paper is organised as follows: Section II summarises the related work in the field and presents the background of this paper. Next, in Section III, we present a use case for measuring and controlling the temperature of a physical room. Based on that we present the Security Cost Modelling Framework and explain the building blocks needed to measure security costs at runtime. Finally, in Section IV we give an outline of future work in the field.

\section{\uppercase{Related Work}}
\label{sec:Related Work}

\noindent There are various approaches, platforms and frameworks supporting the CPS and IoT movement. Derhamy et al. (2015) summarise commercially available IoT frameworks including the IoTivity framework \cite{ref08}, the IPSO Alliance framework \cite{ref09}, the Light Weight Machine to Machine (LWM2M) framework \cite{ref10}, the AllJoyn framework \cite{ref11} and the Smart Energy Profile 2.0 (SEP2.0) \cite{ref12}. Most of the cloud-based frameworks follow a data-driven architecture in which all involved IoT-components are connected to a global cloud using one SOA protocol. The Arrowhead Framework \cite{ref13}, on the contrary, follows an event-driven approach, in which a local cloud is governed through the use of core systems for registering and discovering service, authorisation and orchestration. Since everything within an Arrowhead Local Cloud is a service, new supporting systems can be developed and added to the already existing ones. 

\vspace{-2pt}

Regarding cyber security, there are many studies proposing approaches and frameworks which focus on evaluating security without referring to the resulting costs. Additionally, some of the presented approaches are limited by the usage of a single metric, like process performance in Dumas et al. (2013), and Gruhn and Laue (2006)\nocite{ref14,ref15}. Even though this metric could help to estimate the costs of security it is mainly used to evaluate the process of software implementation. Other related work focuses on methods for measuring how secure a specific system is by evaluating whether a security control has been implemented or not \cite{ref16,ref17,ref18,ref19}. Unfortunately, these approaches provide little insight into how to measure the costs of security. Yee (2013)\nocite{ref20} provides a summary of related work regarding security metrics. He first explains that many security metrics exist, but most of them are ineffective and not meaningful. Furthermore, the author provides a definition of a “good” and a “bad” metric and applies his definition on various frameworks in a literature research.

\vspace{-2pt}

This paper builds on Ivkic et al. (2019)\nocite{ref07} where we introduce an Onion Layer Model for formally describing how the costs of security can be modelled within a CPS. This initial investigation included a mathematical expression for describing the costs of security during the interaction of components and their performed security-related tasks. Additionally, we showed how the Onion Layer Model could be used to evaluate the costs of security for two specific use cases in an exemplary evaluation. To extend this work the key new contribution of this paper is to present an approach for automatically identifying the components of an interaction and their performed tasks at runtime. Furthermore, we extend the previous mathematical expression by transforming it to consider time including a metric catalogue, allowing modelling the costs of security for interactions over a period of time. This allows applying the Onion Layer Model over a longer period of time to be able to   measure, compare and analyse the costs of security of a CPS at runtime. 

\section{\uppercase{Discussion on Modelling Security Costs}}
\label{sec:architecture}

\noindent In this section we present the Security Cost Modelling Framework and its building blocks, which are necessary to measure security costs at runtime. First, we present a use case where a component with a sensor, another component with an actuator and an IoT Framework are interacting with each other. Based on that use case we then propose a framework and discuss how it could be used to measure the security costs at runtime. The proposed framework in Figure 1 includes the Onion Layer Model from our previous work, which uses additional mechanisms in order to identify communicating components and their performed tasks at runtime. In addition to that the framework also uses a Cost Metric Catalogue for measuring the cost of security.

\subsection{Closed-Loop Temperature Control}

\noindent In many respects, the closed-loop control view in Figure 1 corresponds to the most fundamental definition of a CPS. One component ($C_1$) uses a sensor to measure the physical world, while another component ($C_2$) uses this information to change it. Based on that the following use case consists of a component, which uses a temperature sensor to measure a room's temperature ($C_1$), while another component controls an air-conditioning system ($C_2$) to control it. First, $C_1$ becomes part of an existing CPS by registering the temperature sensor to the IoT Framework (step 1). Next, before $C_2$ decides whether the room needs to be cooled down, it sends a request to the IoT Framework asking for a component which is capable of measuring the room's temperature (step 2). However, before the IoT Frameworks returns the endpoint of such a component it verifies whether $C_2$ is authorised for such an interaction (step 3). If it is, the next step is to search the component registries (step 4) and return the temperature sensor component (step 5). After that $C_2$ requests in a loop the room temperature from $C_1$ (step 6), which uses the sensor to measure it and returns the measured value (step 7). Finally, $C_2$ verifies if a limit has been reached (e.g. greater than 25 degrees Celsius) and decides whether to activate the air-conditioning system or not (step 8). Figure 3 shows the sequence diagram including all steps of the described Closed-Loop Temperature Control use case.

\subsection{Security Cost Modeling Framework}

\noindent The Onion Layer Model from from Ivkic et al. (2019), as shown in Figure 2, can be used to describe security costs that occur each time an interaction is executed. In this context, an interaction is defined as a unit of work, which is executed at a specific time, serves a specific purpose and can be treated in an coherent and reliable way independent of other interactions. Furthermore, it includes one or more participating components that perform a number of different tasks. In relation to the use case in Figure 3 the Closed-Loop Temperature Control represents an interaction that involves three components ($C_1$, $C_2$, IoT Framework), which perform a total of seven tasks. 

\noindent \begin{figure}[!h]
	\centering
	{\epsfig{file = 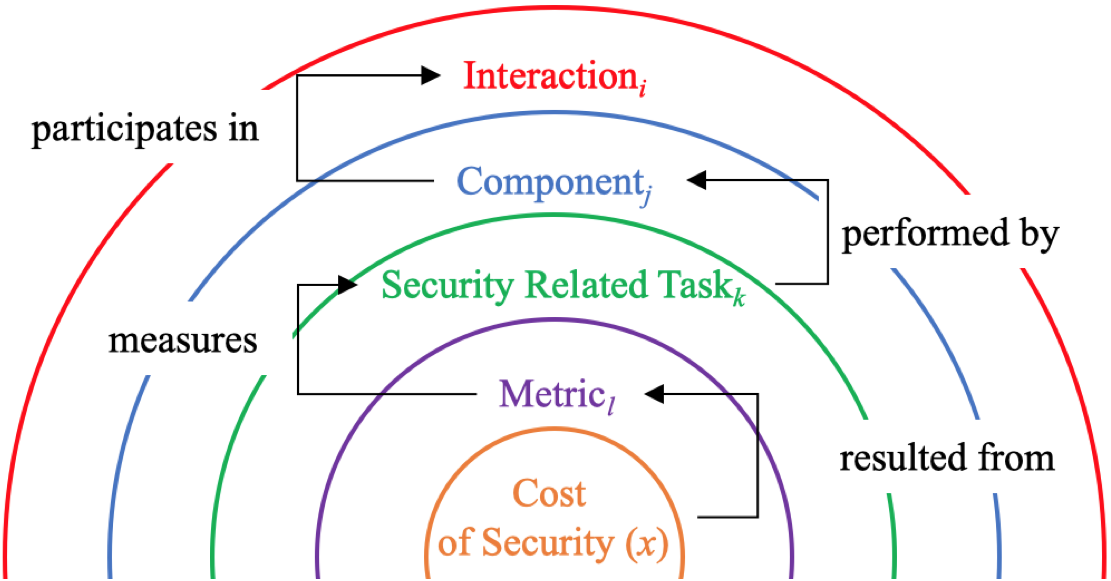, width = \columnwidth}}
	\caption{Onion Layer Model for Modelling Security Costs}
	\label{fig:fig3}
\end{figure}

To measure the security costs of all interactions the Onion Layer Model in Figure 2 suggests to form a sum of sums. The first sum (\textcolor{xred}{$\sum_{i=1}^{n}$}) represents all existing interactions of a CPS, while the second (\textcolor{xblue}{$\sum_{j=1}^{m}$}) summarizes all components within one interaction. The next sum (\textcolor{xgreen}{$\sum_{k=1}^{o}$}) aggregates all security-related tasks which have been performed by a one component. Finally, the last sum (\textcolor{xpurple}{$\sum_{l=1}^{p}$}) adds up all metrics which have been used to measure the performance of a specific security-related task. In our previous work \cite{ref07} the sum of sums has only been used to describe how the cost of security could be modelled within a CPS. 

\begin{equation}
f_{t} = \textcolor{xred}{\sum\limits_{i=1}^{n}} \textcolor{xblue}{\sum\limits_{j=1}^{m}} \textcolor{xgreen}{\sum\limits_{k=1}^{o}} \textcolor{xpurple}{\sum\limits_{l=1}^{p}} \textcolor{xorange}{x}_{t_{\textcolor{xred}{i}\textcolor{xblue}{j}\textcolor{xgreen}{k}\textcolor{xpurple}{l}}}
\end{equation}

Now, to extend this work and to be able to aggregate the security costs at a specific point in time\noindent \begin{figure*}[!h]
	\centering
	{\epsfig{file = 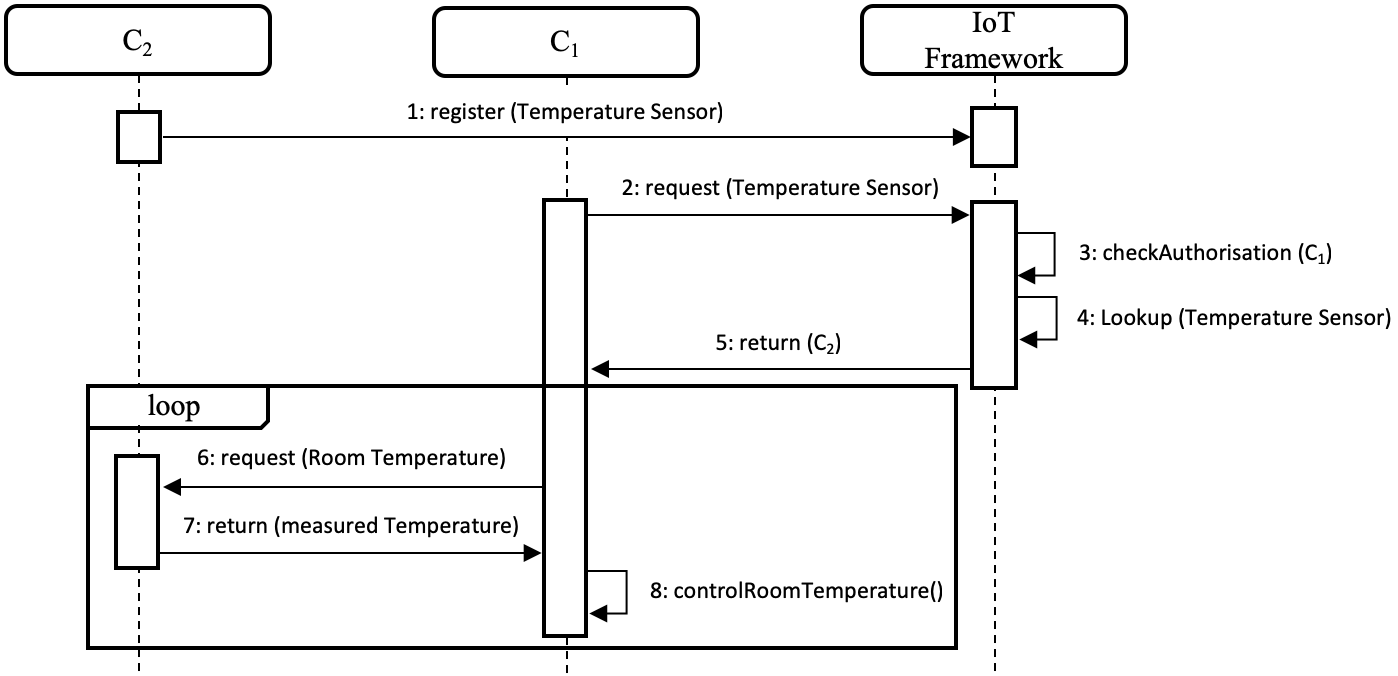, width = \textwidth}}
	\caption{Sequence Diagram for the Closed-Loop Temperature Control Use Case}
	\label{fig:fig2}
\end{figure*} the approach has been extended by a time function \textit{$f_{t}$} as shown in (1). This allows measuring a specific security-related task performed by a component, which participates in an interaction at a specific point in time. Furthermore, measuring the same task periodically allows aggregating the temporal course of security costs for this task.

To measure the security costs of an interaction as they are occurring, the Onion Layer Model needs to know the participating components, the performed tasks and which metrics need to be used. As shown in Figure 1, a Component Monitoring mechanism could listen to the Internet Protocol (IP) address and port of all components. Each time a component sends a message to another one, the mechanism would create a record containing at least the date and time, the sending and receiving endpoints (sender/receiver IP and port). Similar to that a Task Tracing mechanism could log the performed tasks for each component. Additionally, this mechanism should categorize all tasks in use case-related and security-related tasks. Finally, a Cost Metric Catalogue could provide  a set of metric types which can be used to measure the security costs of the previously identified security-related tasks. The combination of the presented mechanisms (Component Monitoring, Task Tracing), the Cost Metric Catalogue and the Onion Layer Model in (1) allows measuring the security costs of interactions at runtime. 

Another interesting aspect of the proposed approach is that the measured runtime data could be used to visualize an interaction, its participating components and their performed tasks. Over time, an interaction with all its components can quickly become incomprehensible, making it difficult for people to keep track of what is going on. To solve this problem the measured runtime data of \textit{"which component communicates with which"} and \textit{"which component performs which tasks and when"} could be used to create a simpler and more comprehensible graph. For instance, Figure 4 shows a possible visual representation of the Closed-Loop Temperature Control interaction including its participating components and their performed tasks:

\noindent \begin{figure}[!h]
	\vspace{-0.7cm}
	\centering
	{\epsfig{file = 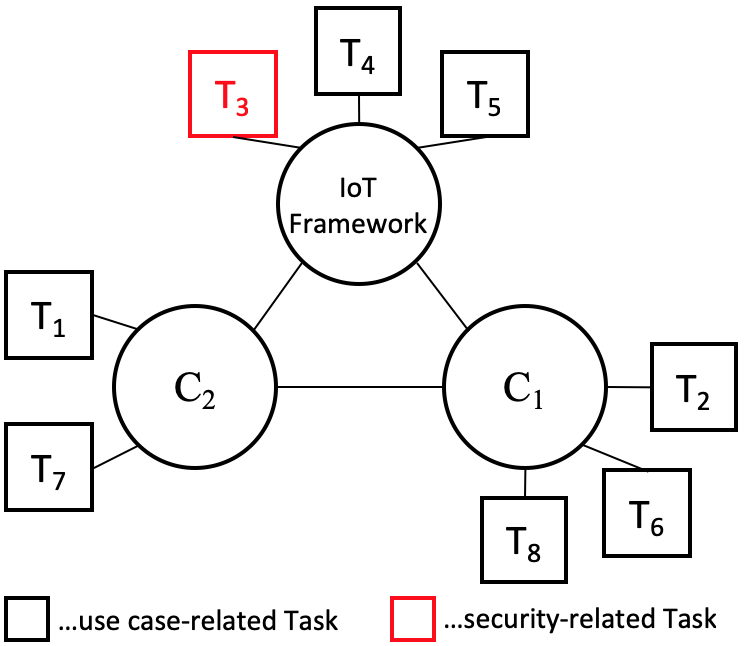, width = \columnwidth}}
	\caption{Visual Representation of an Interaction}
	\label{fig:fig4}
\end{figure} 

This graph enables the visualisation of interactions while they are happening and improves their comprehensibility. Additionally, the graph makes possible the comparison of interactions and identification of performance issues (e.g. bottlenecks).

\subsection{Intended Experimental Design}

\noindent To measure security costs in an experimental study the Closed-Loop Temperature Control use case will be implemented and evaluated. For the implementation we are planning to use a representative IoT Framework which is capable of registering and discovering services and verifying requests for authorisation. The Arrowhead Framework \cite{ref13} could be a possible candidate, since its Core Services (Service Registry, Authorisation and Orchestration) already provide some of the required functionalities. Regarding other requirements (Component Monitoring, Task Tracing) further investigation of the framework is needed to identify whether it already provides the necessary mechanisms, or if they need to be implemented, yet. Furthermore, we plan to execute the Closed-Loop Temperature Control interaction consecutively in a loop of n runs (e.g. n = 50 runs) using two different workloads (WL). In $WL_1$ a representative security messaging protocol (S) will be used to support an encrypted communication between the components ($C_1$ and $C_2$) and the IoT Framework, while $WL_2$ will use an insecure protocol (I). In a first evaluation we are planning to use a representative set of the following four metrics for measuring the performed tasks for each run: 

\begin{itemize}
	\item $M_1$: duration in milliseconds (ms)
	\item $M_2$: Central Processing Unit (CPU)-usage in percent (\%)
	\item $M_3$: Read Access Memory (RAM)-usage in Megabyte (MB)
	\item $M_4$: packet-size of data packages in Kilobyte (KB)
\end{itemize}

The following table summarises the setup of the planned experimental evaluation including the two WLs ($WL_1$, $WL_2$), the number of runs (n) , the used messaging protocols (secure protocol = S, insecure protocol = I) per WL and the metrics for measuring the performed tasks ($M_1$, $M_2$, $M_3$, $M_4$):
\begin{table}[h]
	\caption{Workloads for Experimental Evaluation}
	\begin{tabular}{|c|c|c|l|l|}
		\hline
		\textbf{WL$^{\mathrm{*}}$} & \textbf{Runs} & \textbf{Protocol} & \multicolumn{2}{c|}{\textbf{Metrics}} \\ \hline
		\multirow{4}{*}{$WL_1$} & \multirow{4}{*}{n} & \multirow{4}{*}{S} & \multicolumn{2}{l|}{$M_1$: duration (ms)} \\
		&  &  & \multicolumn{2}{l|}{$M_2$: CPU-usage (\%)} \\
		&  &  & \multicolumn{2}{l|}{$M_3$: RAM-usage (MB)} \\
		&  &  & \multicolumn{2}{l|}{$M_4$: packet-size (KB)} \\ \hline
		\multirow{4}{*}{$WL_2$} & \multirow{4}{*}{n} & \multirow{4}{*}{I} & \multicolumn{2}{l|}{$M_1$: duration (ms)} \\
		&  &  & \multicolumn{2}{l|}{$M_2$: CPU-usage (\%)} \\
		&  &  & \multicolumn{2}{l|}{$M_3$: RAM-usage (MB)} \\
		&  &  & \multicolumn{2}{l|}{$M_4$: packet-size (KB)} \\ \hline
		\multicolumn{2}{l}{$^{\mathrm{*}}$Workloads} 
	\end{tabular}
\end{table}

The runtime information provided by the Component Monitoring and Task Tracing mechanisms at runtime will be used during the experimental evaluation in combination with the Onion Layer Model. The idea is to use the representative metrics to measure each performed task of a component for each run. Then, the following aggregation can be done to measure the costs of using a messaging protocol ($P = \{S, I\})$ for each run ($n = 50$) and each metric ($m = 4$):

\begin{equation}
x_P(i) = \sum\limits_{i=1}^n\sum\limits_{j=1}^{m} M_{j}(i)
\end{equation}

As shown in (2) $x_P(i)$ represents the aggregation of the measured costs of using protocol $P$ for run $i$, while $M_j(i)$ represents the metric $j$ used to measure the costs for each run. Now, as shown in (3) the security costs ($x_{SC}$) can be calculated by the differenc between the two aggregations of using the secure protocol $x_S(i)$ and the insecure protocol $x_I(i)$:

\begin{equation}
x_{SC}(i) = \sum\limits_{i=1}^n x_S(i) - x_I(i)
\end{equation}

\section{\uppercase{Future Work}}
\label{sec:future_work}

\subsection{Implementation \& Evaluation}

\noindent As mentioned in the previous section we will implement the Closed-Loop Temperature Control use case and evaluate its security costs in an experimental study. In this regard we will first investigate a representative IoT Framework, which preferably already includes most of the required functionalities and mechanisms implemented. In addition to that the selected IoT Framework has to be extensible in order to be able to implement missing mechanism and functionalities. Once the use case is implemented we will conduct an experimental study as described in 3.3 using the predefined WLs, protocols (S, I) and representative metrics ($M_1$, $M_2$, $M_3$, $M_4$).

\subsection{Normalisation \& Conversion}

\noindent Even though the presented Security Cost Modelling Framework suggests evaluating security costs at runtime, it implies using metrics with measurement results which can be aggregated. In other words, $M_1$ provides results, which cannot be aggregated with the other metrics. Due to incompatible units, a metric measuring the duration in ms cannot directly be aggregated with another metric measuring the load of a CPU in \%. Another problem is that when using two or more metrics with different units the results may need to be interpreted. For instance, when using all four of the proposed metrics in two runs the measurements might provide the following results: 

\begin{itemize}
	\item $x_1$ = 5 ms + 10 \% + 5 MB + 10 KB
	\item $x_2$ = 10 ms + 5\% + 10 MB + 5 KB
	\item is $x_1 < x_2$ or $x_1 > x_2$
\end{itemize}

Without normalisation of the results it is impossible to tell which of the two measurements is "better" or "cheaper" in terms of security costs. Therefore, when using a metric catalogue in combination with the Security Cost Metric Framework we need a method for either normalising or converting measurement results to a general Cost Unit.

\subsection{Security Costs \& Compliance}

\noindent As already mentioned, monitoring communicating components, tracing their performed tasks and measuring resulting security costs opens up many new possibilities. However, the security costs of e.g. two systems, or two interactions (which serve the same purpose) cannot be directly compared without knowing how secure the system or the interaction is. For instance, if $System_A$ and $System_B$ produce the same security costs for the same tasks they have performed, it does not directly imply that they have the same level of security. $System_A$ might be using a less secure algorithm for encrypting its messages than $System_B$. So, in order to make those two systems comparable in regard to security costs it is also necessary to evaluate how secure both systems are. 

Bicaku et al. (2018b) \nocite{ref21} proposed a Monitoring and Standard Compliance Verification Framework, where they monitor whether a specific security control has been implemented/activated on the target system. Furthermore, they propose  to first extract the security controls from established standards and then provide a mechanism how to monitor if they have been implemented/activated. A combination of the Security Costs Modelling Framework and the Monitoring and Standard Compliance Verification Framework from Bicaku et al. (2018b) could be used to make two systems comparable in regard of security costs and security compliance. We will investigate these two approaches and verify whether it is possible to combine them in future work.

\section{\uppercase{Conclusion}}
\label{sec:conclusion}

\noindent In this paper, we presented a framework, which can be used to measure security costs at runtime. We first presented a close-to-reality use case, which uses an IoT-component to measure the physical world ($C_1$ using a temperature sensor) and another one to affect it ($C_2$ controlling an air-conditioning system). In addition to that an IoT Framework is integrated in this use case, which manages service lookup and authorisation requests. Next, we presented the Security Cost Modelling Framework, which is an extension of our previous work and explain the missing building blocks (Component Monitoring, Task Tracing, Cost Metric Catalogue) to be able to measure the security costs at runtime. Finally, we describe how we intend to evaluate the security costs of the presented use case in an experimental study. This included the design of the experiment, the description of the WLs, runs (n), protocols (S, I) and representative metrics (duration, CPU-usage, RAM-usage, packet-size). Furthermore, we showed how the costs of security will be estimated at runtime by putting all building blocks of the presented Security Cost Modelling Framework together.

The main contribution of this paper is a framework, which can be used to measure security costs at runtime. This Security Cost Modelling Framework will be enhanced by conducting an experimental study as described in Section 3.3 in future work. Furthermore, we will implement the Security Cost Modelling Framework, which uses the outputs of the proposed mechanisms to measure the security costs of the closed-loop temperature control interaction at runtime. Summarising, the main goal is to develop the Security Cost Modelling Framework, which identifies the interacting components and their performed tasks of an interaction at runtime and measures the resulting costs of providing security.

\section*{\uppercase{Acknowledgements}}
\label{sec:acknowledgements}

\noindent Research leading to these results has received funding from the EU ECSEL Joint Undertaking under grant agreement n737459 (project Productive4.0) and from the partners national programs/funding authorities and the project MIT 4.0 (FE02), funded by IWB-EFRE 2014 - 2020 coordinated by Forschung Burgenland GmbH.

\vfill
\bibliographystyle{apalike}
{\small\bibliography{references}}

\vfill

\end{document}